\documentclass[prc,floatfix,nofootinbib,showpacs,showkeys,reprint]{revtex4-1}
\usepackage[latin1]{inputenc}
\usepackage{amsmath}
\usepackage{amsfonts}
\usepackage{amssymb}
\usepackage{graphicx}
\usepackage{url}

\begin{document}

\title{Comment on "Comparison of optical potential for nucleons and $\Delta$ resonances" by Arie Bodek and Tejin Cai }

\author{U. Mosel}
	\email[Contact e-mail: ]{mosel@physik.uni-giessen.de}
	\affiliation{Institut f\"ur Theoretische Physik, Universit\"at Giessen, Giessen, Germany}

\begin{abstract}
In \cite{Bodek:2020wbk} Bodek and Cai have tried to extract the potentials for nucleons and $\Delta$ resonances from electron-nucleus inclusive scattering data. They find that the $\Delta$ potential is considerably more attractive than that of the nucleons. This result is at variance with the results of a multitude of analyses of (e,A) and ($\gamma A$) interactions, performed about 35 years ago.
\end{abstract}

\maketitle

In a recent publication Bodek and Cai \cite{Bodek:2020wbk} have presented potentials for nucleons and $\Delta$ resonances in nuclei \cite{Bodek:2020wbk} obtained from a phenomenological analysis of inclusive electron-nucleus scattering data\footnote{The authors always speak about 'optical potentials' both in their title and their text, but they are concerned only with the real parts of these.}. These potentials are an essential input into event generators for electron and in particular neutrino generators. The nucleon potentials extracted by Bodek and Cai are roughly in line with earlier data obtained from proton-nucleus collisions the potentials. The $\Delta$ potentials that these authors find are always more attractive than the average potential for nucleons. For example, for $^{12}C$ they find $U^\Delta \approx 1.5\,U^{\rm nucleon}$. The purpose of this comment is to point out that this result is in stark contrast to all former analyses of electro-nuclear reactions. 

Potentials for the $\Delta$ resonance in nuclei have been determined in a multitude of theoretical analyses of ($\gamma A$) and (eA) reactions about 30 - 40  years ago. One of the earliest studies was that of Hirata et al \cite{Hirata:1978wp} where a potential depth for the $\Delta$ of - 30 MeV was obtained. Freedman et al. \cite{Freedman:1981dz} also obtained a value for the $\Delta$ potential of - 30 MeV. Later, more refined studies of ($\gamma,A$) and ($e,A$) reactions by Garcia-Reco et al \cite{GarciaRecio:1989xa} obtained a $\Delta$ potential $V_\Delta = -33\, \rho/\rho_0$ MeV where $\rho$ is the $r$-dependent radial coordinate and $\rho_0$ the central density. This is close to the potential obtained in the study of the Delta self-energy in nuclear matter performed in \cite{Oset:1987re} where a value of $V_\Delta = -23 \pm 5\: \rho/\rho_0$ MeV was obtained (all potentials given at zero momentum).  A good summary of the progress achieved both experimentally and theoretically can be found in the book by Ericson and Weise \cite{Ericson:1988gk} which contains detailed theory and reviews of many analyses until then. These authors give a value of -30 MeV for the depth of the $\Delta$ potential in $^{12}C$ and $^{16}O$. Essential for all these results is a careful evaluation of the non-resonant background terms.

All of these $\Delta$ potentials are clearly less binding than the nucleon potentials of about 40 - 50 MeV depth and establish an approximate relation of $U^\Delta \approx 2/3\,U^{\rm nucleon}$ between the two. The result obtained by Bodek and Cai is thus in gross contradiction to all these findings.

\end{document}